\documentclass[12pt,fleqn]{article}
\usepackage{amsmath}
\usepackage{latexsym}
\usepackage{epsfig}
\newlength{\dinwidth}
\newlength{\dinmargin}
\newcommand{\f}[2]{\frac{#1}{#2}}

\begin{document}                   
\def\ra{\rightarrow}
\def\dt{\mbox {\boldmath $\Delta$}}
\def\pmu{p_\mu}
\def\ds{\displaystyle}
\def\tb{\bar{\tau}}
\def\ep{\varepsilon}
\def\pbmu{\bar{p}_{\mu}}
\def\pbnu{\bar{p}_{\nu}}
\def\pb{\bar{p}}
\def\k{{\bf k}}
\def\as{\alpha_s}
\def\q{{\bf q}}
\def\FF{{\cal F}}
\def\ss{\sigma}
\def\ssh{\hat{\sigma}}
\def\bks{\!\!\!\!\!\!\!\!\!}
\def\th{\hat{t}}
\def\uh{\hat{u}}
\def\sh{\hat{s}}
\def\o{\omega}
\def\g{\gamma}
\def\G{\Gamma}
\def\s{\sigma}
\def\de{\partial}
\def\op{\omega_p}
\newcommand{\be}{\begin{equation}}
\newcommand{\ee}{\end{equation}}
\newcommand{\bea}{\begin{align}}
\newcommand{\eea}{\end{align}}
\newcommand{\nn}{\nonumber}
\title{Model (In)dependent Features of the Hard Pomeron}
\author{G. Camici and M. Ciafaloni \\ 
Dipartimento di Fisica, Universit\'a di Firenze \\
and INFN Sezione di Firenze\\
{\em Largo E. Fermi 2, 50125 Firenze}\\
{DFF 260/11/96}}
\date{}
\maketitle
\thispagestyle{empty}

\begin{abstract}
We discuss the small-$x$ behaviour of the next-to-leading BFKL 
equation, depending on various smoothing out procedures of the running 
coupling constant at low momenta.
While scaling violations (with resummed and calculable anomalous dimensions)
turn out to be always consistent with the renormalization group, we argue that
the nature and the location of the so-called hard Pomeron are dependent on the  
smoothing out procedure, and thus really on soft hadronic interactions.
\end{abstract}
\vspace*{4 cm}
\centerline{\hspace*{-1 cm} PACS 12.38.Cy}

\addtocounter{page}{-1}
\newpage

Interest in the high energy behaviour of Quantum Chromodinamics  
\cite{1} has recently revived because of the experimental finding 
\cite{2} of rising structure functions at HERA and has triggered a number of 
papers about the next-to-leading (NL) BFKL equation \cite{3}
and the corresponding anomalous dimensions \cite{4,5}.

One of the interesting features of the NL BFKL equation is supposed to be 
the description of running coupling effects, which on one hand raises the 
question of its consistency with the renormalization group (R. G.) and on 
the other hand emphasizes the problem of the singular transverse momentum 
integration around the Landau pole, sometimes referred to as the IR 
renormalon problem \cite{7}.

The purpose of this note, which is based on a simple treatment of the NL 
BFKL equation, is to emphasize the distinction between the  (leading twist) 
R. G. features, which are genuinely perturbative and thus model independent, 
and the hard Pomeron features which will turn out to be strongly dependent 
on how the effective running coupling is smoothed out or cut off at low  
values of $\k^2=O(\Lambda^2)$.

The problem of the consistency with the R. G. of the BFKL equation with 
running coupling was already analyzed by Collins and Kwiecinski \cite{8}  
by introducing a sharp cut-off in the transverse momentum integrations.
Modifications to the bare Pomeron due to the running coupling were also 
analyzed by Lipatov \cite{8bis}, based on some boundary conditions in the 
soft region which, in our opinion, are eventually equivalent to setting 
a cut-off (see below).
The cut-off dependence of BFKL type equations 
has also been investigated \cite{9}.
Here we whish actually to point out that different ways of {\em smoothing out} 
the running coupling in the large distance, 
small-$\k$ region yield {\em different} answers for the nature, location and 
strength of the (bare) hard Pomeron, while keeping the validity of the leading 
twist renormalization group factorization, as perhaps to be expected.

Let us consider the BFKL equation with running coupling introduced in Refs. 
\cite{8} and \cite{10}
\begin{equation}
f_A(t)=f_{0A}(t)+\f{\bar{\as}(t)}{\o}\int 
d t^\prime K(t,t^\prime)f_A(t^\prime),
\end{equation}
where $f_A(t)=\sqrt{\k^2}\FF_A(\k^2)$ denotes the  unintegrated gluon 
structure function in the hadron $A$, as a function of 
$t=\log(\k^2/\Lambda^2)$, and the BFKL kernel $K(t,t^\prime)$, which 
possibly contains NL contributions, is supposed to be scale invariant with 
the spectral representation 
\begin{equation}
K(t,t^\prime)=\int\f{d\g}{2\pi i}e^{(\g-\f{1}{2})(t-t^\prime)}
\chi(\g)=
\int_{-\infty}^{+\infty}
\f{d\s}{2\pi}e^{i\s(t-t^\prime)}\chi\left(\f{1}{2}+i\s\right)
\end{equation}
which is also assumed to be symmetrical in $t$ and $t^\prime$, 
so that $\chi\left(\f{1}{2}+i\s\right)$ is even in $\s$.

The form (1) of the NL equation was proved to be valid for the $N_f$-dependent 
part of the NL kernel in Ref. \cite{5}. But we have also emphasized 
\cite{11} that the scale of $\as$ can be changed, together with a 
corresponding change in the scale invariant kernel, 
so as to leave the leading twist 
solution invariant, at NL level accuracy.
Therefore, by assuming Eq. (1) we do not emphasize the scale $\k^2$ as the 
natural scale for the running coupling\footnotemark, but we rather consider a
reference form of the BFKL equation which has been widely analyzed previously 
[7-10].
\footnotetext{Actually, the NL calculation\cite{5} suggests that rather 
$\q^2=(\k-\k^\prime)^2$ is the natural scale of the running coupling.}

We shall also assume that the effective coupling $\bar{\as}(t)=
\f{N_c\as(t)}{\pi}$ is smoothed out around the pole at $t=0$, and that the 
inhomogeneous term $f_0(t)$ is peaked at some value $t=t_0>0$. We shall 
often consider in the following the examples
\begin{align}
\as(t)=&\f{1}{bt}\Theta(t-\bar{t})+\f{1}{b\bar{t}}\Theta(\bar{t}-t),
\quad\quad\quad(\bar{t}>0),\\
f_0(t)=&\delta(t-t_0),
\end{align}
but our discussion will not be limited to these particular forms.

Let us start noticing that if $\as(t)\leq\as(t_M)$ has a maximum at 
$t=t_M$, and $\chi(\f{1}{2}+i\s)\geq\chi(\f{1}{2})$ (as is the case 
for both the leading and the NL expressions considered so far \cite{3,5}),
the Pomeron singularity $\op$ has the upper bound 
\begin{equation}
\op\leq\bar{\as}(t_M)\chi\left(\f{1}{2}\right).
\end{equation}

This follows \cite{8} from general bounds  
on the norm of $K$, and needs no further 
explanations. We shall show, however, that besides this general result, 
the properties of the Pomeron singularity are very much dependent on the 
model for $\as(t)$.

In order to understand this point, we shall discuss the solution to Eq. (1) 
by using a quasi-local approximation of the kernel $K$ valid around the 
end-point of the $\g$-spectrum in Eq. (2), i.e.,
\begin{equation}
K(t,t^\prime)\simeq \chi\left(\f{1}{2}\right)\left(
1+a^2\de_t^2+....\right)\delta(t-t^\prime),\quad\quad
a^2=\f{1}{2}\f{\chi^{\prime\prime}\left(\f{1}{2}\right)}{\chi
\left(\f{1}{2}\right)}.
\end{equation}

This equation is based on the simplest polynomial expansion of $\chi(\g)$ 
around $\g=1/2$, which of course has already been used in the literature 
\cite{8bis}.
Our purpose here is to obtain, by means of the expansion in Eq. (5) a simple
picture of the hard Pomeron properties, and to show that this picture is 
stable when some higher order polynomial approximation is used.

By replacing Eq. (6) into Eq. (1), the latter becomes an inhomogeneous 
second order differential equation in the $t$ variable, of the form
\begin{equation}
\f{\as(\bar{t})}{\as(t)}\left(f(t)-f_0(t)\right)=
\f{\bar{\o}}{\o}\left(1+a^2\de_t^2\right)f(t),
\end{equation}
\begin{equation}
\left(\bar{\o}\equiv\bar{\as}(\bar{t})\chi(\f{1}{2})\right),\nn
\end{equation}
and its homogeneous part is just a Schroedinger-type equation with a 
given potential $V$ and wave number $k$, given by the expressions
\begin{align}
V(t)=&\f{1}{a^2}\f{\o}{\bar{\o}}\left(\f{\as(\bar{t})}{\as(t)}-1\right),
%\f{2}{\chi^{\prime\prime}(\f{1}{2})}\f{\o}{\bar{\as}(t)},\\
\\
k^2=&\left(1-\f{\o}{\bar{\o}}\right)\f{1}{a^2}.
\end{align}

While this potential is always linear for $t>\bar{t}$ because of the 
perturbative behaviour of the running coupling, its form may vary considerably
according to how $\as(t)$ is smoothed out in $-\infty<t<\bar{t}$ 
(Fig. 1 (a), (b)) or cut off (Fig. 1 (c)).

In the case $\as$ is sharply cut-off (Fig. 1 (c)), one has an infinite potential well, 
which has, of course, a discrete spectrum, whose ground state provides the 
Pomeron pole, in agreement with previous analyses [7-9].

If instead the smoothed out coupling has a flat behaviour below $\bar{t}$ 
(Fig. 1 (a)) the spectrum is continuum and the Pomeron is a branch 
cut singularity with branch point
\begin{equation}
\op=\bar{\as}(-\infty)\chi\left(\f{1}{2}\right).
\end{equation}

In this case the eigenfunctions are just plane waves for $t\ra-\infty$, and 
there appears to be no reason why their phase-shift should be fixed by some 
condition \cite{8bis}, contrary to the case with cut-off.

Finally in the intermediate case of Fig. 1(b) there may be an 
isolated singularity too, depending on the depth of the well.

Whatever the relevant case is, the Green's function of the corresponding 
Schroedinger equation is easily calculable, and provides the solution 
of Eq. (7) if we set $f_0(t)=g_A\delta(t-t_0)$.
A straightforward analysis shows that for $t>t_0$ such solution takes 
the factorized form
\begin{equation}
\sqrt{\k^2}\FF_A(\k^2,Q_0^2)=f_A(t,\o)=
f_R(t,\o)g_At_0f_L(t_0,\o),\quad\quad \left(t>t_0=\log\f{Q_0^2}{\Lambda^2}
\right)
\end{equation}
where $f_R$($f_L$) denote the regular solution of the homogeneous 
equation for $t\ra\infty$ ($t\ra-\infty$). Their explicit form for 
$t>\bar{t}$ is
\begin{align}
f_R(t,\o)=&f_+(t,\o)\\
f_L(t,\o)=&f_-(t,\o)+R(\o)f_+(t,\o)
\end{align}
where $R(\o)$ is the reflection coefficient of the well, 
$f_+$($f_-$) denote the regular (irregular) solutions for $t\ra\infty$ in 
the linear potential, given by the expressions 
\begin{align}
f_\pm\equiv&\int_{C_\pm}\f{d\g}{\sqrt{2\pi}i}e^{\left(\g-\f{1}{2}\right)t-
\f{X(\g)}{\bar{b}\o}},\quad\quad (t>\bar{t}),\\
X(\g)\equiv&\int_\f{1}{2}^\g\chi(\g^\prime)d\g^\prime=
\chi\left(\f{1}{2}\right)\left(\g-\f{1}{2}\right)+
\f{1}{6}\chi^{\prime\prime}\left(\f{1}{2}\right)\left(\g-\f{1}{2}\right)^3
+....\quad,
\end{align}
and $C_+$ ($C_-$) denote the regular (irregular) contours for the Airy 
functions (Fig. 2).

The reflection coefficient (or $S$-matrix) $R(\o)$ in Eq. (13)
is easily found - starting from $k^2=-\chi^2<0$, $(\o>\bar{\o})$ - for the 
simple model of $\as(t)$ in Eq. (3). In such case the wave functions are 
just exponentials for $t<\bar{t}$, and by the customary matching 
procedure we find
\begin{equation}
R(\o)=\f{\chi-L_-(-\chi^2)}{L_+(-\chi^2)-\chi},
\end{equation}
where $L_+$ ($L_-$) denotes the logarithmic derivative at $\bar{t}$ of 
$f_+$ ($f_-$) in Eq. (14). 

Therefore, the coefficient $R(\o)$ contains 
the Pomeron singularity of the wave number $\chi$, which in this case is 
a continuum starting at $\o=\bar{\o}$.
It is also clear that $R(\o)$ will contain the Pomeron singularity 
for each of the models described in Fig. 1. In particular there may be 
an isolated pole due to the possible vanishing of the denominator in 
Eq. (16).

Our final result is thus that for large $t$, the unintegrated gluon 
density takes the factorized form
\begin{equation}
\sqrt{\k^2}\FF_A(\k^2,Q_)^2)=
f_+(\o,t)g_At_0\left[f_-(\o,t_0)+R(\o)f_+(\o,t_0)\right]\quad
(t>t_0).
\end{equation}

The $t$-dependent factor in Eq. (17), quoted in Eq. (14), is just the 
na$\ddot{\imath}$ve 
regular solution \cite{8,10} of the homogeneous equation corresponding to
Eq. (1), for which only the perturbative form of $\as(t)$ matters.
Its large $t$, small $\o$ behaviour in the anomalous dimension regime 
\begin{equation}
\bar{b}\o t>\chi\left(\f{1}{2}\right)
\end{equation}
is dominated by a saddle point which reproduces the R. G. behaviour with 
running coupling as follows
\begin{equation}
f_+(\o,t)\simeq\f{1}{\sqrt{\k^2}}\f{1}{\sqrt{-\chi^\prime(\g_L)}}
\exp\int^t\g_L\left(\as(t^\prime)\right)dt^\prime,\nn\tag{19a}
\end{equation}
\begin{equation}
1=\f{\bar{\as}(t)}{\o}\chi(\g_L),\nn\tag{19b}
\end{equation}
\addtocounter{equation}{1}
where $\g_L$ is the perturbative branch of the anomalous dimension.

The $t_0$-dependent factor consists instead of two terms. 
One is the na$\ddot{\imath}$ve perturbative term $f_-(\o,t_0)$, which is 
the irregular 
solution of Eq. (1) and only has, like $f_+$, an essential singularity at 
$\o=0$. 
The other one contains instead the hard Pomeron singularity, due to the 
$R(\o)$ factor, which mixes the irregular with the regular solution. 
It appears, therefore, that for large $t_0=\log(Q_0^2/\Lambda^2)$, 
the Pomeron term is suppressed with respect to the na$\ddot{\imath}$ve 
term by 
inverse powers of $Q_0^2/\Lambda^2$, just because $R(\o)$ 
multiplies the {\em regular} solution. This explains the mechanism by 
which the perturbative evolution picture emerges in the case of two large 
scales, i.e. $Q^2\gg Q_0^2\gg\Lambda^2$, as often assumed in the literature
 \cite{12}.

The above transparent picture of the bare hard Pomeron properties emerging 
from the simplest quasi-local approximation of $K$ has the obvious defect of 
incorporating only leading twist solutions in Eq. (19b), due to the use
of a quadratic approximation for $\chi(\g)$.

Let us show that the above picture is only slightly modified when higher 
twist solutions of Eq. (19b) are introduced. We can devise, for instance, a 
polynomial approximation to $\chi(\g)$ of order $2(n+1)$ which has (i) the 
quadratic part fixed as before and (ii) $n$ higher twist solutions of the 
saddle point equation
\begin{equation}
\o\bar{\as}^{-1}(t)=\chi^{(n)}(\g_n),
\end{equation}
fixed also, together with their mirror solutions $\tilde{\g}_n=1-\g_n$, where 
$\g_0=\g_L(\as(t))$ denotes the perturbative branch.

The Green's function of the corresponding $2n+2$-order 
differential equation will then 
contain $n+1$ combinations of type (11), as follows
\begin{equation}
\sqrt{\k^2}\FF_A(\k^2,Q_0^2)=\sum_{i=0}^n f_R^{(i)}(t,\o)
g_L^{(i)}(t_0,\o),\quad\quad (t>t_0),
\end{equation}
where the functions $f_R^{(i)}(t)$ denote the regular solutions for 
$t\ra\infty$, depending on the $i^{th}$ branch of the anomalous dimension, 
while the functions $g_L^{(i)}(t_0)$ are to be determined from the 
continuity (discontinuity) requirements of the Green's function and of its 
first $2n$ derivatives ($2n+1-th$ derivative).

It is not difficult to see that the $g_L$'s are the ratio of two determinants 
of Wronskian type involving the $f_R^{(i)}$'s and the $f_L^{(i)}$'s. 
For large $t_0$, they are asymptotically given by 
\begin{equation}
g_L^{(i)}(t_o)\sim\left[f_R^{(i)}(t_o)\right]^{-1},\quad\quad (t_o\gg 1).
\end{equation}

We see from Eqs. (21) and (22) that the gluon distribution now contains higher 
twist terms, as expected. Nevertheless, the properties of the leading twist 
contribution are unchanged.
In fact, $f_R^{(0)}$ has the behaviour 
quoted in Eq. (19) and $g_R^{(0)}$ has the 
decoupling property in Eq. (22) for large enough $t_0$. Of course the 
detailed hard Pomeron properties determined by the mixing of 
$f_+$'s and $f_-$'s in the $f_L$'s will be more complicated and depending on 
various scattering coefficients.

The above argument shows that, in general, the structure of the gluon 
distribution is correctly hinted at by the polynomial approximations
to $\chi(\g)$, and leads us, therefore, to the following conclusions.

Firstly, whatever the weight of the Pomeron is, the solution of Eq. (1) 
satisfies the expected leading twist R. G. factorization property, for any 
value of $t_0$, and precisely
\begin{equation}
\k^2\FF(\k^2,Q_)^2)=\f{1}{\sqrt{-\chi^\prime(\g_L)}}
\exp\left(\int^t\g_L(t^\prime)dt^\prime \right)K_L(\o,t_0),
\quad\quad\quad\quad \Big(\bar{b}\o t>\chi({1}/{2})\Big),
\end{equation}
with
\begin{equation}
K_L(\o,t_0)=\sqrt{\bar{b}\o}t_o\left[f_-(\o,t_0)+R(\o)f_+(\o,t_0)
+....\right].
\end{equation}
The $t$-dependent factor in Eq.(23) is the one predicted by the R.G. , with an
additional coefficient, which is relevant for NL calculations \cite{5,11}.

Secondly, the hard Pomeron coupled to hadrons
is really a nonperturbative phenomenon, which is 
very much dependent on the behaviour of the strong coupling
in the soft region $k^2 = O(\Lambda^2)$. Only if the initial scale is
large enough does the Pomeron decouple, and the BFKL evolution becomes 
genuinely perturbative.      

If this is the case, then exploring the Pomeron structure and unitarity 
corrections to it becomes really a strong interactions problem in which soft 
physics plays a major role, without much distinction between short and 
long distance contributions.

On the other hand, the $t$-dependent Pomeron singularity present in Eq. 
(19) will still play a role, as a singularity of the anomalous dimension
expansion. Therefore, if the variable $\as(t)\log(1/x)$ is not too large, 
the standard perturbative approach will still be applicable, 
provided a resummation to all orders is performed, along the lines reported 
elsewhere \cite{11}.

\begin{center}
{\bf Acknowledgements}
\end{center}

We are grateful to Bryan Webber and Jan Kwiecinski for interesting discussions,
and to the CERN theory division for hospitality, while part of this work was 
being done. This paper is supported in part by M.U.R.S.T. (Italy) and by 
E.C. contract\# CHRX-CT93-0357.

\newpage

\begin{figure}[htb]
%\vspace*{-0.8 cm}
\centerline{\psfig{figure=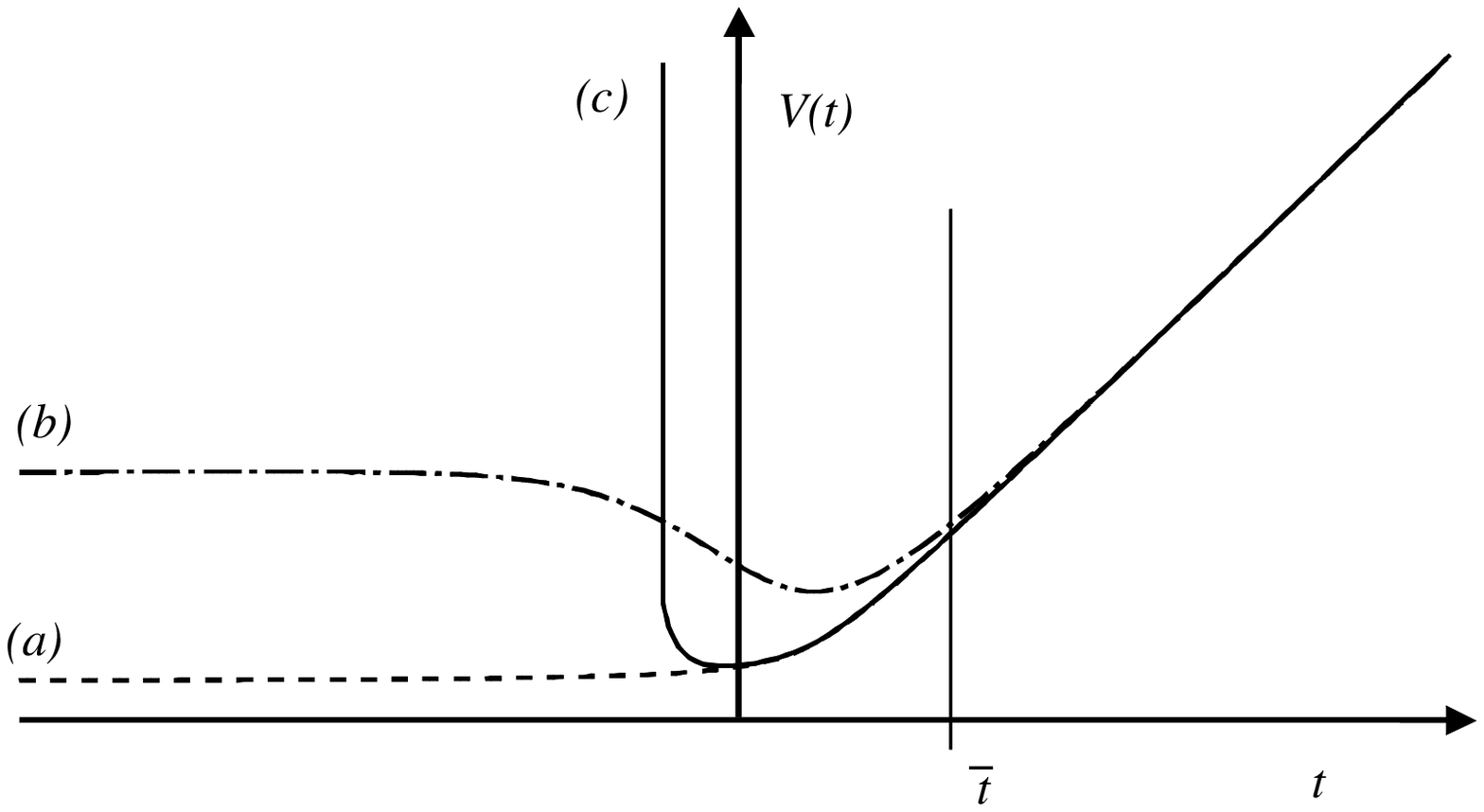}}
\vspace*{-16 cm}
\caption{Form of the potential $V\sim \as^{-1}$ in Eq. (8), if 
the behaviour of $\as(t)$ for $t <\bar{t}$ is (a) flat, (b) with a maximum
and (c) with a sharp cut-off.}
\end{figure}

\newpage 

\begin{figure}[htb]
\vspace*{-4 cm}
\centerline{\psfig{figure=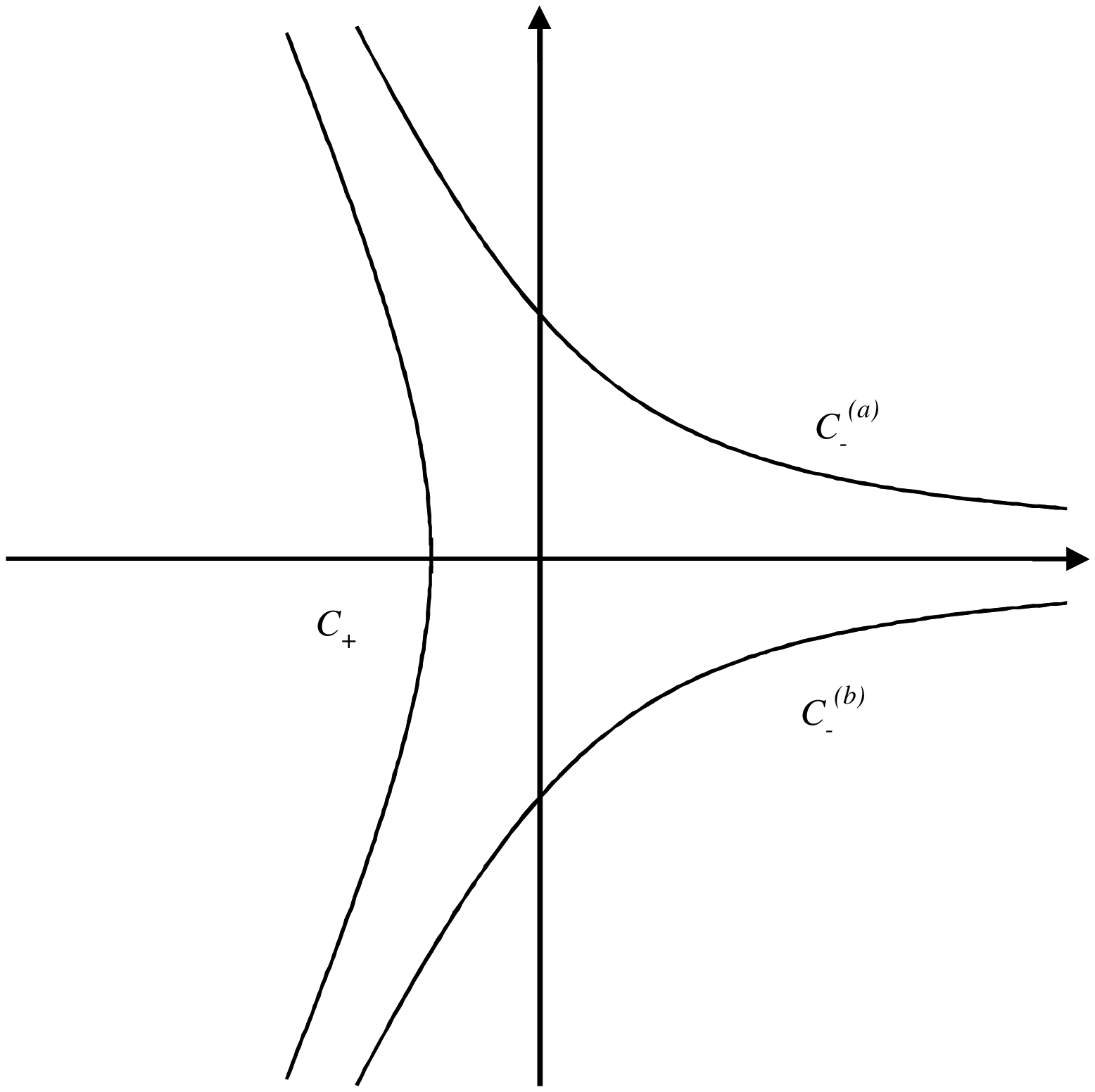}}
\vspace*{-8 cm}
\caption{Contours $C_+$($C_-$) for the regular (irregular) solutions
in the linear potential. The contour $C_-$ is meant to be the average 
over $C_-^{(a)}$ and $C_-^{(b)}$.}
\end{figure}


\begin{thebibliography}{99}
\bibitem{1} L.N. Lipatov, Sov. J. Nucl. Phys. 23 (1976) 338;\\ 
E.A. Kuraev, L. N. Lipatov and V. S. Fadin Sov. Phys. JETP 45 (1977) 199;\\ 
Ya. Balitskii and L. N. Lipatov, Sov. J. Nucl. Phys. 28 (1978) 822.
\bibitem{2} H1 Collaboration, T. Ahmed et al., Nucl. Phys. B 439 (1995) 471;
Preprint Desy 96-039;\\
ZEUS Collaboration, M. Derrick et al., Z. Phys. C 65 (1995) 379; 
Preprint Desy 95-193.
\bibitem{3} V. S. Fadin and L. N. Lipatov, Yad. Fiz. 50 (1989) 1141;
Nucl. Phys. B406(1993)259; Nucl. Phys. B 477 (1996) 767.\\
V. S. Fadin and R. Fiore, Phys. Lett. B 294 (1992) 286;\\
V. S. Fadin, R. Fiore and A. Quartarolo, Phys. Rev. D 50 (1994) 5893;\\
V. S. Fadin, R. Fiore and M. I. Kotsky, Phys. Lett. 
B 359 (1995) 181; HEP-PH 9605357;\\
V. del Duca Phys. Rev. D 54 (1996) 989; Phys. Rev. D (1996) 4474;
\bibitem{4} S. Catani and F. Hautmann, Phys. Lett. B 315 (1993) 475; 
Nucl. Phys. B 427 (1994) 475.
\bibitem{5} G. Camici and M. Ciafaloni, Phys. Lett. B 386 (1996) 341.
\bibitem{7} See, e. g., G. P. Korchemsky and G. Sterman Nucl. Phys. 
B 437 (1995) 415;\\
E. Levin Nucl. Phys. B 451 (1995) 207;\\
M. Beneke and V. M. Braun Nucl. Phys. B 454 (1995) 253;\\
Yu. L. Dokshitzer, G. Marchesini and B. R. Webber, 
Nucl. Phys. B 469 (1996) 93. 
\bibitem{8} J. Kwiecinski, Z. Phys. C 29 (1985) 561;\\
J. C. Collins and J. Kwiecinski, Nucl. Phys. B 316 (1989) 307.
\bibitem{8bis} L. N. Lipatov, Sov. Phys. JETP 63(1986)904.
\bibitem{9} J. C. Collins and P. V. Landshoff, Phys. Lett. B 276 (1992) 196;\\
M. F. McDermott, J. R. Forshaw, G. G. Ross,Phys. Lett. B 349 (1995) 189;\\
J. Bartels, H. Lotter and M. Vogt, HEP-PH 9511399;\\
K. D. Anderson, D. A. Ross and M. G. Sotiropoulos, HEP-PH 9602275.
\bibitem{10} L. V. Gribov, E. M. Levin and M. G. Ryskin,
Phys. Rep. 100 (1983) 1.
\bibitem{11} G. Camici and M. Ciafaloni, to appear.
\bibitem{12} A. H. Mueller and H. Navelet, Nucl. Phys. B 282 (1987) 727.
\end{thebibliography}
\end{document}